# REFLECTIVITY STUDIES AND PRODUCTION OF NEW FLAT MIRRORS FOR THE CHERENKOV THRESHOLD DETECTORS AT CERN

J. Buesa Orgaz*, M. van Dijk, D. Banerjee, J. Bernhard, M. Brugger, N. Charitonidis,
A. Ebn Rahmoun, M. Lazzaroni, V. Marchand, I. Ortega Ruiz, E.G. Parozzi, G. Romagnoli,
F. Sanchez Galan, T. Schneider, J. Tan, M. van Stenis, CERN, Geneva, Switzerland

*Abstract*

Cherenkov threshold detectors (XCET) are used for identifying particles in the experimental areas at CERN. These detectors observe Cherenkov light emitted by charged particles travelling inside a pressurized gas vessel. A key component of the XCET detector is the 45-degree flat mirror reflecting the Cherenkov light towards the photomultiplier (PMT). A thorough analysis and optimization was conducted on the design and materials of this mirror, along with the surface coatings and coating techniques. A suitable manufacturing process was selected, and the first mirror prototype was produced, installed, and tested in the East Area at CERN. Experimental data obtained during beam tests is presented to assess the efficiency of the new coating and materials used.

## INTRODUCTION

Cherenkov threshold detectors (XCET) [1] have been used for identifying secondary charged particles in CERN's experimental areas since the 1970s. A schematic overview of the functional components of the XCET can be seen in Figure 1. The 45-degree flat mirror holds particular importance, as it reflects Cherenkov light [2, 3] towards the PMT. The primary objectives for the redesign of this mirror were to minimize the mirror thickness, the scattering, and the energy loss of the incoming beam [4] while maximizing reflectivity, and to find a manufacturing processes that could produce cost-effective yet high-quality mirrors.

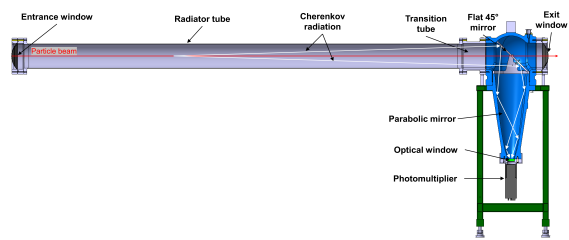

Figure 1: XCET detector design (CERN East Area).

## MATERIALS AND METHODS

### Current Design Limitations

The current mirror consists of a 50 µm reflective aluminium foil fixed on an ellipsoidal frame positioned at a 45-degree angle. Its installation history is uncertain, and it has likely degraded since its initial installation. The next generation uses polymer films with reduced density compared to aluminium, decreasing material on the beam axis. Key considerations when choosing materials for the mirror included high radiation resistance, optimal adhesion to the aluminium frame, uniform coating adhesion [5], and low surface roughness. Eight polymer films from 4 different families were studied: Polyethylene Terephthalate (PET), Polyimide (PI), Polyethylene Naphthalate (PEN) and Polystyrene (PS). Three foils were sourced from Goodfellow (PET, PI, PS), while the others included Mylar A (PET), Kapton 200 HN (PI), Kapton 300 HPP-ST (PI), Melinex S (PEN), and Kaladex 2000 (PEN).

### Coating of Polymer Samples

To evaluate the reflectivity of the polymer foils under identical conditions, the 8 polymers were simultaneously coated using electron-beam physical vapor deposition (EB-PVD) from Laybold Optics [6]. A 100 nm layer of aluminium was applied for enhanced UV reflectivity followed by a 20 nm protective layer of $MgF_2$. The deposition rate of the aluminium was 0.4 nm/s and the $MgF_2$ was 0.8 nm/s.

### Novel Coating Method

The Thin Film Lab at CERN has developed a novel thermal coating method that enhances UV reflectivity on the 45-degree mirror. Aluminium, the best material providing UV reflectance, absorbs in the far ultraviolet due to oxidation. To mitigate this, a specially designed recipe has been developed. Aluminium (Al) is pre-melted on Tungsten spirals and evaporated using thermal PVD. The 100 nm Al layer is applied in a flash mode in just a second. Immediately after, a 20 nm layer of $MgF_2$ is added to protect against oxidation while also creating selective enhancement in the far UV. This recipe was used on a 50 µm Mylar type A film.

### Test Beam Validation

The old mirror and the new Mylar mirror with the novel coating were deployed in XCET043 of the T10 beamline in the CERN East Experimental Area of the Proton Synchrotron complex [7]. The test beam consisted of electrons, pions, kaons and protons, in varying proportion as a function of momentum. For the negative beam, the proton content is very low (<0.2%). At -10 GeV/c, the beam consists almost completely of pions (>90%), the rest being kaons, muons, and a small fraction of electrons. At -4.5 GeV/c, the beam is around 15% electrons, 80% pions, the rest being kaons and muons. At these settings the Cherenkov thresholds of pions in $CO_2$ are 1.07 bar at 4.5 GeV/c and 0.2 bar at 10 GeV/c. The beam had a 1% momentum spread.

---
* jan.buesa.orgaz@cern.ch

A LeCroy WaveRunner 104MXi-B oscilloscope was used to record the analog signal of the PMT (ET Enterprises 9814QB). The coincidence of two scintillators triggered the signal acquisition, one upstream and one downstream of the XCET. This ensured that each particle passed through the full length of the radiator. The signal was integrated over a window of 40 ns, starting around 10 ns before the onset of the pulse. The light generation should scale linearly with the pressure above the threshold pressure for a given particle at a given momentum, and lighter particles will generate more light. Any particle below its Cherenkov threshold will not generate any light. The signals recorded below threshold serve as the pedestal, i.e. the charge observed in the PMT when there is a trigger but no Cherenkov light. Care was taken to stay below the kaon threshold for all data analyzed, so there is always a small but significant pedestal to be found.

## RESULTS

### Optical Measurements

Total reflectivity analyses were conducted on the 8 polymer samples coated via EB-PVD, the novel flash-coated Mylar mirror and the old mirror. Total reflectivity values are shown in Figure 2, employing a fixed 30-degree Angle of Incidence (AoI) to ensure consistency with past measurements. An increase of up to 2% is anticipated at an AoI of 45 degrees from previous measurements.

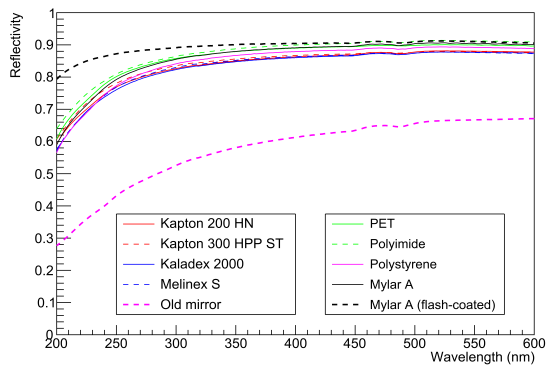

Figure 2: Total reflectivity measurements at 30-degree AoI of the 8 polymer samples by EB-PVD, the old mirror and the new 45-degree mirror by flash coating. Measured using a Lambda 650 UV/Vis Spectrometer from Perkin Elmer with the Universal Reflectance Accessory.

A simulated Cherenkov light spectrum was generated using a fixed refractive index for $CO_2$ at 1 bar, $n$=1.00045 [3]. This light yield was convoluted with the measured reflectivity of the mirrors, shown in Figure 3. In the spectral range considered, detectable light increased by 41% using the new mirror.

All optical components still have appreciable efficiency down to at least 180 nm so the absolute increase should be higher than the estimated 41%. It was measured in situ at the T10 beamline.

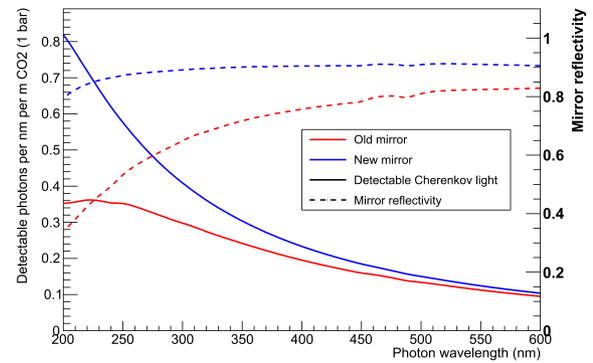

Figure 3: Mirror reflectivity (dashed line) and quantity of detectable Cherenkov light (solid line), for the old (red) and new (blue) mirrors. The detectable Cherenkov light was calculated by convoluting a simulated Cherenkov spectrum per nm per m of $CO_2$ at 1 bar with the mirror reflectivity.

### Beam Test Measurements

The integrated charge was recorded for at least 30k particles for each pressure and momentum. At -10 GeV/c the electron fraction is negligible so the pion peak was the only distinguishable feature. At -4.5 GeV/c the electron peak was fitted, and the pion peak was fitted when they cross the threshold. The pedestal is fitted when clearly distinguishable and otherwise assumed to be fixed. This was found to be a good approximation when comparing datasets. The values taken from the fits are the means of the fitted Gaussians. The pedestal was then subtracted from the charge observed. The results are shown in Figures 4 and 5.

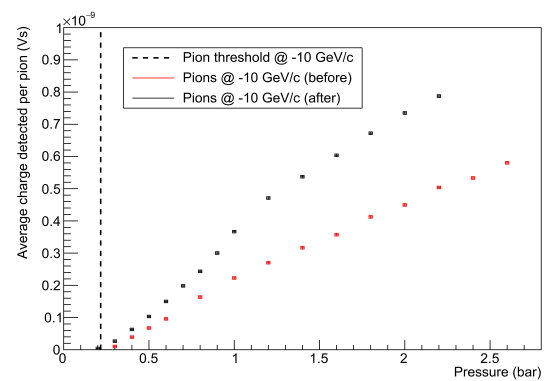

Figure 4: Mean of Gaussian fit to pion peak of charge spectrum at -10 GeV/c, before and after the mirror replacement.

To compare the absolute improvement of the XCET mirror quality, the ratio of the situation after and before was calculated and shown in Figure 6.

## DISCUSSION

While the overall reflectivity for all EB-PVD polymer films changes significantly from 200 nm to 300 nm, there was no significant difference between the reflectivity of the various polymer films, less than 5% over the full range. The observed variations among polymer samples coated with EB-PVD might be related to a variety of factors, including surface roughness and coating adherence, which influence the uniformity and quality of the reflective coating.

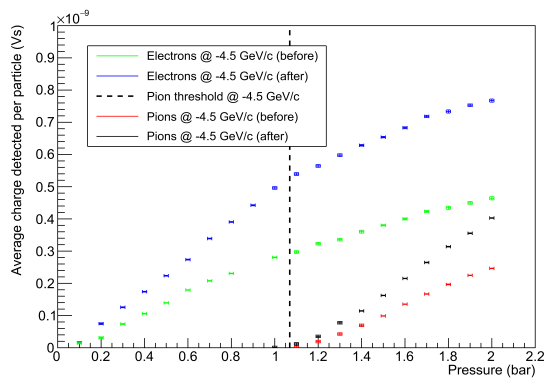

Figure 5: Mean of Gaussian fit to electron/pion charge spectrum at -4.5 GeV/c, before/after the mirror replacement.

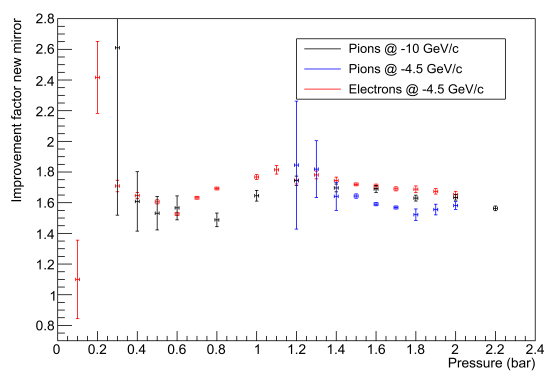

Figure 6: Improvement factor of the new mirror, calculated as the after/before ratio for the -10 and -4.5 GeV/c datasets.

The thermal flash coating approach significantly improved UV reflectivity when compared to EB-PVD, notably in the 200-300 nm region. At a wavelength of 200 nm, the flash coating showed a remarkable improvement of 20% in reflectivity, with a reflectivity of 79% compared to EB-PVD of 59%.

The primary difference in the two approaches is the significantly different coating deposition rates. Despite the fact that both procedures take place in a high-vacuum environment, oxidation still occurs, and high-energy photons are particularly susceptible to its effects [8] . Therefore, optimizing the deposition rate of the Al and $MgF_2$ layers is imperative, along with minimizing the interval time between layer deposition. In the case of EB-PVD, the aluminium layer required 250 seconds for full deposition, followed by an additional 180 seconds until the $MgF_2$ layer was deposited. In contrast, the flash coating method achieved aluminium evaporation within a a second, immediately followed by $MgF_2$ evaporation within 15 seconds.

The ratio of the observed charge before and after the exchange of the mirror, shown in Figure 6, shows the substantial improvement in the quality of the mirror. The larger error bars stem from the errors being approximately constant for all measurements, hence dominating in the ratio for small values of the total observed charge. Over most of the range, excepting those with the lowest signal, the increase is around 60%, substantially more than the expected increase of 41% over the 200-600 nm range. This implies there is an additional contribution coming from outside of this range.

The change in the light intensity observed in Figures 4 and 5 showcase the expected increase in light generation with pressure. A change in slope is observed at around 1 bar or so. Two possible causes are considered: PMT saturation and absorption of the light on its path. The inner diameter of the radiator is limiting. As the pressure climbs, the Cherenkov angle grows and eventually the light reflects off the inside of the radiator tube. This surface is electropolished stainless steel is not so reflective, in the UV in particular [9]. This surface becomes a factor when the Cherenkov angle reaches 27.2 mrad. The onset of the kink, beyond 1 bar or so, matches this reasonably. For electrons, the Cherenkov angle at 1 bar $CO_2$ is 30 mrad. As this feature occurs at the same pressure for both the old and the new mirror with different total light intensities, it is attributed to the limited reflectivity of the radiator inner surface, and not PMT saturation.

## CONCLUSION

It was found that the choice of coating technique significantly outweighs the impact of polymer substrate selection on total reflectivity. While diffusive reflectivity was not quantified, further studies on reflectivity and surface roughness could enhance understanding of substrate influences. Mylar A exhibited favourable compatibility with the frame-glue interface and coating adhesion. The novel thermal flash coating technique emerges as the most effective method employed. With all manufacturing processes conducted in-house at CERN, production is rapid, unit costs are reduced compared to industry standards, and reflectivity results are notably improved. The newly fabricated mirror using the novel coating method was deployed and tested in the T10.XCET043 in the secondary beamline of the CERN Proton Synchrotron East Experimental Area. Comparing the performance of the old and new mirror showed a consistent improvement of around 60% in detected light intensity. Future tests with the same mirror will give insight on the impact of aging. The goal to establish and verify a fabrication process for such mirrors by means of in situ testing was therefore met.